\documentclass[a4,12pt]{article}
\usepackage{graphicx}
\usepackage{psfig}
\usepackage[dvips]{epsfig}
\begin{document}




\begin{titlepage}

\begin{flushright}

physics/0406012

\end{flushright}

\vspace{.5cm}

\begin{center}

{\Large \bf Quantum Entanglement, Cognition $\&$ The Processes of
Inference }

\vspace{1cm}

{\large R. Chakraborty{\footnote{mitrarc@yahoo.com}},$^a$\, I.
Mitra{\footnote{indranil@theory.saha.ernet.in}},$^b$\,}
\vspace{5mm}

{\em $^a$ Rashmandir Vidyalaya,\\
 BK -61,Salt Lake City, Kolkata-91}\\
{\em $^b$Department of Physics, New Alipore College\\
L Block, Kolkata 700 053, India}\\
\vspace{.5cm}
\end{center}

\vspace{.5cm}

\centerline{{\bf{Abstract}}}

\vspace{.5cm}

\begin{small}
Church's hypothesis and Godel's theorem may provide constraints on
mental processes.As a relief quantum entanglement may lead to a
definite proposal as regards the nature of reality and how much of
it we are able to know and how do we know it.We deal with these
questions and have devised a Gedanken Experiment showing the
redundancy of principle of reality. We thereby propose a model
describing the processes of making conclusions from a given set of
premises though without the reference of neurobiological
processes at present.We argue thereafter that the physical laws
of nature may not be universal as it seems, being dependent on
some physical processes namely computation  and the states of
neural cognitive states which may be governed yet by some other
rule.

\end{small}

\end{titlepage}

\newpage





\section{Introduction}

The purpose of this paper is to investigate in general terms the
observed phenomena,nature of reality and the way we reach
conclusions from a set of premises.For the development of any
scientific discovery the initial set of patterns account for most
of the observations and experiments which are fundamental to its
realizations.Furthur development can be achieved by the
construction of elaborate equipment,techniques,esoteric vocabulary
and skills. The scientific method encompasses both the
mathematical and empirical sciences, and we briefly discuss why it
is necessity that drives science and determines what we mean by
the scientific method. By necessity we mean laws of nature and of
human thought that scientific theories have to try and explain. By
the statement that science is driven by necessity, we mean that
the discipline of science as such addresses aspects of reality
that are governed by necessity. Scientific research is open ended
and exploratory, and finding new and unexplained phenomena is the
main objective. Mathematical sciences - which include mathematics,
computer science, simulations and so on - consist of purely
symbolic structures that are freely produced by the human mind. Of
course all of language consists of signs and words as well, in
that the word is not a thing but can only signify to what it
represents. A mathematical theorem necessarily follows from the
axioms of the subject. It does not depend on the subjectivity of
any mathematician, since all mathematicians will be able to
reproduce the same (universal) result. Evidence in mathematics
consists of the derivation of the result. There are instances when
there are plausible theorems for which no such evidence is
available, and such theorems are called conjectures. Only when a
proof is offered is the matter resolved. Mathematics is that
component of human thought that is entirely and completely
determined by necessity. Mathematical theorems are unconditionally
valid - in all circumstances and for all time. Empirical sciences
- which include physics, chemistry, biology, astronomy, geology
and so on - are theories and explanations of nature based on
empirical verification. Such verification entails producing
evidence - for or against the theory - in the form of data which
is presented to sense perception, and is external to the mind.
Scientific data cannot be something that is produced by the human
mind. Once empirical evidence - namely, data - has been procured,
the mind can then analyze such data to interpret, verify or
falsify various scientific theories.

 The methodology of science, it's so called
mode of inquiry, is determined by necessity.The most powerful
aspect of the sciences is its ability to predict how nature will
behave in the future, or in new and unknown circumstances.
Technology is rooted on devices and instruments that are made
based on scientific principles, which can then successfully
manipulate nature.  Science can even predict the existence of
phenomena that were hitherto unknown. For example, the mathematics
of quantum physics predicted in 1929 \cite{dirac} that an electron
would have an anti-particle as its counterpart. Anti-particles had
never even been thought of until then, let alone having been
detected. And sure enough, when guided by this theoretical
prediction, anti-electrons were experimentally found in 1935.
There are numerous such examples in science.  The question
naturally arises, are there mathematical structures essential for
theory that do not have any measurable manifestation? Richard
Feynman has the following to say. ``It is not true that we can
pursue science completely by using only those concepts which are
directly subject to experiment. In quantum mechanics itself there
is a probability amplitude, there is a potential, and there are
many constructs that we cannot directly measure.''\cite{schwabl}
Are there theoretical constructs in physical theories that have
absolutely no physical consequence, but are nevertheless required
by the mathematical structure of the theory? One can only make
some conjectures. In quantum mechanics, the answer to this
question, in principle, seems to be yes. The predictive power of
science is completely based on the universality and
``objectivity'' of mathematics. The result of mathematical
calculations yield new symbols and structures that are then
identified, by a leap of imagination, with physical objects and
processes. As we proceed we will be tackling in detail the terms
"imagination" etc.  The route essentially depends on a suitable
process of computation.\cite{nielsen} So as a whole we see if we
can link the discovery of physical theories to be synthetic
truths, then the process of computation may be the crucial
component for the theory of knowledge.

As it can very well be understood that at this stage of
development the analysis will be mainly speculative and we will be
considering a mechanism from a purely kinematical point of view.
The question merely rests on the age old question raised by Kant
as to the efficacy of pure mathematics and its methodology. The
answer by pure empiricists that our mathematical knowledge is
derived by induction from particular instances seems to be
inadequate as the validity of of the inductive hypothesis itself
cannot be proved by induction. As we all would possibly agree that
mathematical knowledge is general. It seems strange that that we
should be able to know some truths in advance about particular
things of which we yet have no experience. Again the observable
world around us which are in general our experiences based on
perceptions which gives rise to knowledge which is particular. But
the amazing fact which has remained is that the application of
mathematics to the observable world has produced general
statements regarding it. We are avoiding at present more general
questions regarding the status of statements and their relation to
the world of universals. Now it is a obvious observation as we
have already stated that the scientific theories do heavily rely
on mathematics.The mode of operation of mathematics and its
progress has been on the other hand entirely dependent on the
process of computation.

At this point we are led into several viewpoints as regards its
functionality. The so called proponents of strong AI
 \cite{minsky} which claim that any sort of thinking is
computation. Apart from dealing with the questions of awareness
and consciousness which we don't intend to address at present (we
would mention some of its implication on our theory in the
conclusions {\footnote{A good reference for all these ideas may be
in \cite{conc}}}) we have a feeling that essentially any
conceivable idea of reaching conclusions is apparently through a
set of rules and methods which is essentially a computation.
Though as has been proposed by Penrose \cite{penrose} that
appropriate physical action of the brain may not be simulated
computationally. We are not yet sure about the proposal yet, the
Godel \cite{godel} Turing arguments as regards the incompleteness
of mathematical reasoning does'not quite guarantee that some sort
of process which is completely non computational may make it
complete.May be incompleteness is the truth itself. Penrose's
hypothesis regarding quantum gravity making a route to solve the
riddles seems an attractive proposal nonetheless seems to be
driven by a sole motivation to bind the nature into a single
whole.

 One of the important aspects of scientific research is to ask
reasonable questions. In the context of analysing physical
theories the fundamental objects under scrutiny is the
meausurement procedure and algorithm themselves. So as a matter
of fact whether some computational process is stoppable or not as
has been advocated in \cite{stapp} is mainly a technical problem.
Analysing meausurement processes immediately leads us to the
world of quantum mechanics. As we all know the paths of quantum
mechanics tries to see nature in terms of of the observer and the
observed the construction of physical theories is inevitably
linked to both the segments as we will look in due course. The
next object under analysis is the corresponding computational
process and algorithms. Surprisingly devising algorithms will
bear new fruits rather than generating them.

 The paper is
organized as follows. In Section 2 we review the Church's Thesis
and Godel Turing arguments applied to the mental processes. In
Section 3 we will discuss a short review of Quantum entanglement
and the basic idea which we will require in the last section-the
central theme of the paper. Section 4 deals in devising a
Gedanken Experiment demonstrating nature of reality and it's
relation to our Observed world. The last section is concerned
with a proposal of a model as regards the description of the
physical universe and the way we reach various conclusions, which
in turn give rise to physical theories.

\section{Church Turing hypothesis $\&$ Godel's Theorem}

Whenever we are into the arena of discussing theories of
processing information we are inevitably led to discuss Church's
thesis.We mainly follow \cite{manin},\cite{suppes}.

\subsection{Church's Thesis :}
(a) A function f is semicomputable iff it is partial recursive.\\
(b) A function  f is computable {\footnote{$x \in {Z^{+}}^m $as
input gives  f(x) as output if $x \in D(f)$ or 0 otherwise.
Computational functions are semi computable}}iff both f and
$\chi_{D(f)}${\footnote{Characteristic function of $X\subset Y
\Rightarrow $$\chi_{X}: Y\rightarrow Z^{+} $ s.t
 $\chi_{X}(x) = 1 $ if $x \in X $ and 2 otherwise.}} are
partial recursive.\\
\vspace{1cm}
\textit{Definition :}\\
(a)  A sequence of partial functions $f_{1}, f_{2}....f_{n}$is
called partial recursive description of the function
\hspace{6cm}$f_{N} = f$\\
if

 \begin{center}
  $\left\{
\begin{array}{cc}
f_{1} ,  & \mbox {belongs to family of basic functions;}\\
f_{i} ,  & \mbox {$i\geq 2$ either belong to family of basic
functions or obtained by recursion}
\end{array}\right\}$
\end{center}

 (b) A function is called partial recursive if it admits a
partial recursive description.

\vspace{.5cm}
Church's thesis is used for
a definition of algorithmic undecidability. Suppose we have a
countable sequence of mathematical problems $ P_{1}, P_{2}.....$.
Furthermore suppose that such a problem has a yes or no answer
and that the conditions on $P_{n}$ are written effectively as a
function of n. We associate to such a problem a mapping,

 \hspace{5cm} $f : Z^{+} \longrightarrow
Z^{+}$\\

\hspace{5cm}$D(f) = \{i  \in  Z^{+} \mid P_{i}$ has yes for an answer \}\\

\hspace{5cm} $ f(i) = 1   \hspace{1cm} if \hspace{1cm}   i \in  D(f)$\\

The problem P is called algorithmically decidable if the functions
f and $\chi_{D(f)}$ are partial recursive. Otherwise P is called
algorithmically undecidable. Church's thesis alows us to solve
such equations in two stages: (1) finding an informal solution
using any intuitive algorithms we need. (2) formalising the
solution. The Church-Turing thesis concerns the notion of an
effective or mechanical method in logic and mathematics.  A
method, or procedure, M, for achieving some desired result is
called 'effective' or 'mechanical' just in case M is set out in
terms of a finite number of exact instructions,M will, if carried
out without error, produce the desired result in a finite number
of steps; M can be carried out by a human being unaided by any
machinery. M demands no insight or ingenuity on the part of the
human being carrying it out. Statements that there is an effective
method for achieving such-and-such a result are commonly expressed
by saying that there is an effective method for obtaining the
values of such-and-such a mathematical function. For example, that
there is an effective method for determining whether or not any
given formula of the propositional calculus is a tautology e.g.
the truth table method is expressed in function-speak by saying
that there is an effective method for obtaining the values of a
function, call it T, whose domain is the set of formulae of the
propositional calculus and whose value for any given formula x,
written T(x), is 1 or 0 according to whether x is, or is not, a
tautology.

\subsection{Turing's Hypothesis :}
 The notion of an effective method is an informal one, and
attempts to characterise effectiveness, such as the above, lack
rigour, for the key requirement that the method demand no insight
or ingenuity is left unexplicated. One of Turing's achievements in
his paper of 1936 \cite{tur} was to present a formally exact
predicate with which the informal predicate 'can be calculated by
means of an effective method' may be replaced. Church did the
same (1936a)\cite{chur} . The replacement predicates that Turing
and Church proposed were, on the face of it, very different from
one another, but they turned out to be equivalent, in the sense
that each picks out the same set of mathematical functions. The
Church-Turing thesis is the assertion that this set contains
every function whose values can be obtained by a method
satisfying the above conditions.

The formal concept proposed by Turing is that of computability by
Turing machine. He argued that whenever there is an effective
method for obtaining the values of a mathematical function, the
function can be computed by a Turing machine. The converse claim
is easily established, for a Turing machine program is itself a
specification of an effective method: without exercising any
ingenuity or insight, a human being can work through the
instructions in the program and carry out the required operations.
If Turing's thesis is correct, then talk about the existence and
non-existence of effective methods can be replaced throughout
mathematics and logic by talk about the existence or non-existence
of Turing machine programs.

\subsection{Godel's Incompleteness Theorem :}

Let T $\subset $ S(A) be the set of true expressions where A is
the finite or countable alphabet and let S(A) be the set of
expressions in A. D$\subset$ S(A) is the set of provable or
deducible expressions.\\
The incompleteness principles of Godel are enumerated as follows:
(i) The set D is enumerable. (ii) The set T is not enumerable if
the semantics of truth is rich enough to include elementary
arithmetic.

So as we see Godel's Theorem can be succinctly be stated as:\\
All formal theories of mathematics satisfy the above incomplete
principles of (i) $\&$ (ii). Therefore if a theory be sufficiently
rich, it always contains true expressions which are not provable.

\vspace{.1cm}

The above methodology may be now turned on to see how the process
may be applied to mental processes. The conclusion of Godel's
argument implies that there are allowed brain process P that can
never stop in spite of the fact that no system operating according
to the rules by which the brain operates could ever reach the
conclusion that P can never stop. As a matter of fact that there
exists a k such that $C(k,k)$ stops iff  $C(k,k)$ can never stop
is knowable cannot be proved within the fintistic type of model
described above.The finite type computer X which can in principle
simulate the mechanical activity of the human brain can be
imbedded in a computer Y whose rules of operation included
implementation of the concept for $\forall$'n', and to which the
Godel/Turing argument can be applied. Then the mind M that could
comprehend both the rules of operation of C and also the logic of
the Godel/Turing proof would be able to compute a value k such
that the proposition that ''$ C(k,k) $ stops iff $ C(k,k) $ can
never stop is knowable",where knowable means information by virtue
of the outputs of Y. The mind M knows that $P_{k}$ is true can
know also that X is knowable entails that X is true,and can
therefore conclude that $ C(k,k)$ can never stop. Thus M can know
more than what is knowable. This is the analog of Godel's theorem
and is not a contradiction. On the other hand the human
mathematician can know only the output of X. He will be able to
reason on the basis of what hypothetical, M is able to know, that
there exists some k, such that $C(k,k)$ can never stop. However it
has not been proved that the only way that he could know this is
by virtue of the stopping of $C(k,k)$. The stopping of $C(k,k)$
may be the unique process which gives the strong result that
$C(k,k)$ can never stop is knowable.

\section{Quantum Entanglement, A Short review:}
One of the key issue in the foundations of quantum mechanics for
all these years has been the Einstein-Podolsky-Rosen argument
\cite{ein} regarding the incompleteness of a quantum theory though
mainly the
Copenhagen interpretation.The basic argument can be stated as\\
\textit{Every element of the physical reality must have a
counterpart in the physical theory}\\
The proposal has been analysed from various contexts and the
notable among them been the Bohmian mechanics,\cite{bohm} Bell's
Theorem  \cite{bell}  and Gleason's Theorem \cite{glea}. We will
not go into the details here but mainly discuss the most notable
feature which the argument has led us into, the idea of Quantum
Entanglement. As we will see in Section 5 our main construction
is very much related to the argument and the idea of entangled
states. Our discussion mainly is on the same lines with
\cite{mackay}

A simple description of the essential non-classical nature of
quantum entanglement is given in Entangled Choices. For a more
detailed and technical analysis, consider the quantum spin of an
electron (or any spin-1/2 particle), which is always found to have
either the value +h/2 or -h/2 in the direction of measurement,
regardless of the direction we choose to measure. Thus an electron
manifests one of only two possible spin states, which we may call
spin up and spin down. It's convenient to represent these states
as orthogonal unit vectors In general an electron's spin state,
Y, at any instant can be represented by a linear combination of
those two possible observable states. The choice of a measurement
direction is equivalent to choosing a basis for expressing the
spin components of the state Y. For any particular
basis we can express the state in the form\\

\hspace{6cm}$|\Psi\rangle  =  c_{1}|+\rangle  +  c_{2}|-\rangle $

where $c_{1} $and $c_{2}$ are complex constants. These constants
encode the probability that the chosen measurement will yield
either result. The probability is simply the norm of the
respective coefficient.We have\\

\hspace{6cm}$c_{1}^{\star}c_{1} + c_{2}^{\star}c_{2} = 1 $

In quantum mechanics, each possible measurement basis is
associated with an operator whose eigenvalues are the possible
results of the respective measurement. For a given xyz basis of
orthogonal space coordinates we can represent the three principle
measurements (i.e., measurements along the three axes) by the
 Pauli spin matrices $\sigma_{x},\sigma_{y},\sigma_{z}$.
The eigenvalues of the measurement operator corresponding to
whichever measurement direction we choose determine the
coefficients $c_{1}$ and $c_{2}$, which represent the
probabilities of the possible outcomes. To see how this works,
suppose $|\theta \rangle$ is the initial state vector of the
electron, and we decide to perform a spin measurement
corresponding to a particular operator . The result is given by
applying to $|\theta \rangle$, using ordinary matrix
multiplication, to give the new state vector $|\beta \rangle$ as
shown below:\\

\hspace{6cm}\textbf{U}$|\theta \rangle =  |\beta \rangle$

If  $|\beta \rangle$ is either pure spin up or pure spin
down(basis states) in the direction of measurement represented by
, it follows that a subsequent measurement in the same direction
must yield the same
result, so $|\beta \rangle$ must be such that\\

\hspace{6cm}$\textbf{U}|\beta \rangle =  q |\beta \rangle$ \\

for some constant q(since state vectors are equivalent up to
length). Thus the constant q is an eigenvalue of the measurement
operator,and $|\beta \rangle$is the corresponding eigen-vector.

Since the arbitrary initial state  $|\theta \rangle$is not in
general an eigenvector of \textbf{U}, it can yield either of the
eigenvectors of the basis states. This reveals the probabilistic
aspect of quantum mechanics.

The eigenvectors constitute a basis for the space of possible
state vectors, so  $|\theta \rangle$can be expressed as a linear
combination of those eigenvectors. If we let  $|\beta \rangle$
$\&$ $|\delta \rangle$ denote the eigenvectors, then we can
express $|\theta
\rangle$as\\

\hspace{5cm}$ c_{1}|\beta \rangle  +  c_{2}|\delta \rangle =
|\theta
\rangle$\\

Again, the norm of each complex coefficient gives the probability
that applied to $|\theta \rangle$ will lead to the respective
eigenstate.

The same analysis can be extended to three or more principle
direction spin operators as  presented above. Each pair of
eigenvectors constitutes a basis for the state space, so we can
express the electron's state vector as a linear combination of the
basis vectors for the desired measurement, and the coefficients
give the probability of that measurement yielding either spin up
or spin down. We can think of these probabilities as the
projections of the initial state vector onto the orthogonal axes
of the chosen measurement basis. Of course, we aren't restricted
to measurements along one of the principle axes. We can measure
the spin of the electron along any spatial axis, and each such
measurement is represented by an operator. We also note that these
directions are purely relative to the state of the particle in
question.

These quantum mechanical predictions have some remarkable
implications.  An experiment can be performed by examining the
spins of two spin-1/2 particles emitted in opposite directions
from the decay of a singlet state with zero total spin. In such a
case conservation of angular momentum requires that the spin state
vectors of the individual particles are precisely opposite, so if
we measure the spin of one of these particles along a certain
direction and find spin up in that direction, then the other
particle must have pure spin down in that direction. Thus, by
measuring the spin of one particle and reducing its state vector
to one of the eigenvectors of the chosen measurement basis, we
automatically collapse the wavefunction of the other particle
onto this same basis.

At first sight this process may not seem very mysterious, since
it's easy to imagine that the two coupled particles are
programmed with opposite spins, such that they will always give
opposite results when measured along any given direction.
However, a detailed  analysis of the quantum mechanical
predictions for all possible combinations of measurement angles
reveals the need for a profound change in the classical view of
the world.

If the measurement of one particle along a fixed direction (in the
xz plane) yields spin down, then the other particle is purely
spin up in that direction. Consequently if we perform a
measurement on the other particle along a direction at an angle of
$\alpha$ from the first measurement, we will see that the
probability of spin up is $\frac{1}{2}cos(\alpha/2)$ and the
probability of spin down in $\frac{1}{2}sin(\alpha/2)$. In a
similar way we can show that if the measurement of the first
particle yields spin up, then the other particle is purely spin
down along that direction, and a measurement of that other
particle along a direction at an angle $\alpha$ relative to the
first will yield spin up with probability
$\frac{1}{2}sin(\alpha/2)$and spin down with probability
$\frac{1}{2}cos(\alpha/2)$.

Hence the probability that the measurements of these two particles
at angles differing by $\alpha$ will both give the same result
(both up or both down) is $\frac{1}{2}sin(\alpha/2)$ , and the
probability that they will yield opposite results (one up and one
down) is $\frac{1}{2}cos(\alpha/2)$ . The angle $\alpha$between
the two measurements can be expressed as x - y  where x is the
angle of the measurement performed on one of the particles and y
is the angle of the measurement performed on the other. The two
particles emitted from a singlet state are said to be entangled,
because regardless of how far apart they travel before the spin
measurements are made, the joint results will exhibit these joint
probabilities.

From a classical standpoint we would imagine that each particle
emerges from the singlet state with, in effect, a set of
programmed instructions for what spin to exhibit at each possible
angle of measurement, or at least what the probability of each
result should be. The usual approach to analyzing these phenomena
classically is to stipulate that a particle's programmed
instructions for responding to a measurement at a given angle
must be definite and unambiguous (rather than probabilistic)
because we classically regard the two measurement angles as
independent, which implies that the measurement on the other
particle could be at the same angle as our measurement of this
particle, and the particles must give opposite results in that
case. Likewise the measurement on the other particle could be 180
degrees away from our measurement of this particle, and the
particles must give equal results in that case. Of course, the
individual measurements can each be either spin up or spin down
in both of these cases, so in principle they could still be
probabilistic tendencies, but classically we have no way of
ensuring perfect correlation (or perfect anti-correlation) of the
joint results of spacelike separated events other than by
definitely programming the spins of each particle for each
possible measurement angle.

From this assumption it follows that the instructions to one
particle are just an inverted copy of the instructions to the
coupled particle. In other words, for each measurement angle from
0 to 2p the programmed response to a spin measurement for one
particle is the opposite of the programmed response of the other
particle at that angle. Furthermore, since we have perfect
correlation if our measurements are at angles that differ by 180
degrees, it follows that the programmed instructions for each
particle are individually anti-symmetric at a phase angle of 180
degrees. For example, if a particle's programmed response for a
measurement at angle a is spin up, then the programmed response
of that same particle for a measurement at angle x + p must be
spin down. Hence we can fully specify the instructions to both
particles by simply specifying the instructions to one of the
particles for measurement angles ranging from 0 to p .

For the sake of completeness it will be worthwhile to state in
brief some of the resolutions of the problem. Numerous discussions
of quantum entanglement discuss hidden variables,\cite{hidd}
\cite{koch}and they derive inequalities involving the correlations
as functions of these so called variables. In a sense, these
variables are redundant, because the essence of the so-called
local realistic premise is that each particle emerges from the
singlet state with definite instructions for the spin it will
exhibit for each possible measurement angle. This implies that
the only relevant free variable is the reference orientation,
since the instruction profile can be oriented in any direction
relative to the measurements with uniformly distributed
probability. Integrating the product of spins over the entire
angular range with a fixed difference angle must yield the
correlation of the two spin measurements.

It might be argued that we need not assume any single particle
exhibits the quantum mechanical probabilities, because these
probabilities can only be evaluated by performing multiple
measurements, so we might imagine that it's necessary to evaluate
the results over a sequence of particle pairs, rather than just a
single pair. Moreover, we might think that this sequence could
possess memory,\cite{mem} making the results of successive pairs
dependent. However, in principle a large number of particle pairs
can be prepared in an identical way, in spacelike-separated
locations, and perform measurements on the pairs independently.
According to quantum mechanics we would still expect the combined
results to satisfy the same correlations. This implies that each
particle pair must embody the overall propensities that we expect
to find manifested in multiple trials.
\section{A Gedanken Experiment}


\begin{figure}[h]
\begin{center}
\includegraphics[]{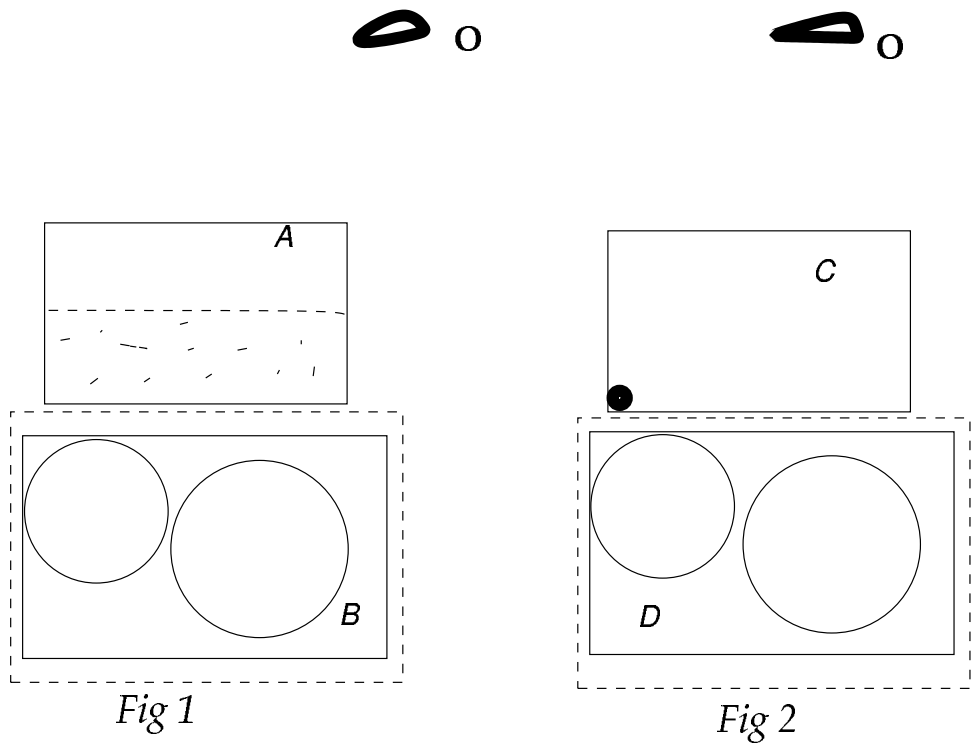}
\end{center}
\caption{Observed $\&$ Real Worlds } \label{}
\end{figure}

The figures depicted above is a diagram for the simple thought
experiment which we describe below.In both Fig A and Fig B Part Y
are identical,as shown enclosed within a dashed box are to be
taken inaccessible to any observer which we term as reality in a
sense that the phenomena of the observable part X and X' of  world
will be dependent on them.So X $\&$ X'are the only parts which the
observers are aware of. So part Y of both the diagrams do not
exist in relation to the observer. The experiment is designed as
follows.\\
 Let us first describe Part Y of the diagrams. As we have
stated earlier part Y is a hypothetical entity which are non
existent to any observer. It consists of an enclosure which
contain two massive balls, which may move or collide in the
enclosure producing observable effects on part X $\&$ X' of the
systems. Some readers may be sceptical about the dynamics of the
motion of the balls etc, but we are not interested in it as the
observer will
only observe Part A.\\

In Fig A the part X of the diagram describes a simple container
containing a liquid. As it is clear from the previous
consideration the effect of part Y on part X is to produce some
disturbance in the liquid and the motion of the surface of the
liquid,inside the container. But the observer being unaware of the
part B of the apparatus has to interpret the motion. One of the
possible explanation of the motion may be given by the assumption
of an existence of a density function $ \rho(x,y,z)$inside the
liquid. To describe a definite model to explain the effective
motion of the liquid it may be assumed a non uniform distribution
of particles along the length of the enclosure as to say \\

\hspace{6cm}$ \rho(x,y,z) = \frac{1}{z}$\\

This implies that there are large number of the particles at the
surface of liquid which continually decrease as one goes along the
liquid upwards till the liquid surface. So the large number of
particles at the bottom may influence some pressure and thereby
force on the next layer of particles which continually goes on to
produce an appreciable motion of the liquid surface.\\
So essentially the success of the model essentially depends on a
suitable choice of the density function which will correctly
describe  the motion of the surface of the liquid as has been
observed.\\

In Fig B Part X' of the diagram denotes again a simple box which
has a charged particle placed on the floor of the box. That the
particle is a charged one has been tested previously in a lab by
suitable electrical experiments(say). Now again as in the previous
case the Part Y of the figure will influence the Part X' and will
in turn effect the motion of the charged particle. So as it is
clear by now that the observer will have to describe the motion of
the charged particle., though the observer is absolutely unaware
of Part Y. The charged particle may be undergoing arbitrary
motions and the observer accounts for it by ascribing an existence
of an electric field $ E(x,y,z)$ inside the container. The correct
description of the charged particle will eventually depend on the
suitable choice of the electric field.\\

So now the two gedanken experiments give rise to some interesting
consequences. First of all it is not necessary that the observed
world may match with the real world. The most important fact
related to this is that a single real world may give rise to many
observable situations and thereby to many descriptions of the
observable world. and the same reality. Now at this stage a
question may be asked that so what the big deal is this, as in any
case we cannot observe the real world and may be cannot really
aspire to know about it why not rename the observable world as the
real one. Well as far naming is concerned it may not be a big
problem, but can we sure whether it will not induce any effects on
the observable world, perhaps not.This is the line of attack we
will take up in the next section where we argue that the real
world though absolutely unobservable and inaccessible yet may be
crucial as far development of physical theories and thereby
knowledge is concerned. We will be giving an example to
illucidate our point.

\section{ A proposed model for the mechanism of generating
physical theories:}

The basic aim of this section is not the the content of knowledge
or it's one of the manifestations, physical theories. On the
contrary we are interested in finding whether there are any
schemes about which we infer conclusions or to be specific enough
to go on to discover physical theories, mathematical theorems etc.
The Godel Turing argument analysed in Section 2 clearly shows that
computational processes may be insufficient to bear all fruits.
Penrose claims that there may be noncomputational \cite{pen}
\cite{noncomp1} \cite{compu}process linked with brain activities.
The model we want to describe here though satisfies the Godel's
incompleteness theorem, yet we invoke some different methods
which do not need that the brain activities need to be
noncomputational in principle. the arguments are furthur enhanced
by the Gedanken experiment performed in Section 4.

So in order to describe the model, the observed world is
essentially demarcated from the real world which is completely
inaccesible to the observer. The only role the real world has is
in producing observable effects to the observed world as we have
seen in the thought experiment described in section 4.

The information content corresponding to the observed world is
encaspulated in the semantics, mathematical tools, symbols and
logical $\&$ semantic connectives. \cite{logic}\\
Let \textsl{a} be the set of alphabets and the rule of syntax,
\textsl{m} corresponds to the sets of mathematical symbols and the
rules of inference, \textsl{q} denotes the set of logical
and semantic connectives.\\
Now,\\
\begin{eqnarray}
\textit{L} &=&  \{s(l)| l\in \textsl{a}, s(l) \equiv statements
\hspace{0.1cm} and  \hspace{0.1cm}propositions
\hspace{0.1cm}\hspace{0.1cm}of  linguistics\}\\
  \textit{M} &=&  \{s(m)|m\in \textsl{m}, s(m) \equiv  statements
  \hspace{0.1cm}  and \hspace{0.1cm} propositions \hspace{0.1cm}
   of   \hspace{0.1cm}mathematics\}\\
  \textit{C} &=&   \{s(c_{l},c_{m})|c_{l},c_{m}\in
  \textsl{q}, s(c_{l},c_{m})
  \equiv  statements \hspace{0.1cm} of \hspace{0.1cm}
  \hspace{0.1cm} the \hspace{0.1cm} connectives\}
\end{eqnarray}

So as we can all understand that the full information of the
observable world is encoded in the above constructed sets. To be
more specific any information regarding the observable world is
nothing but a statement or proposition obeying the formal rules of
mathematics and linguistics.\\
So if we construct the direct product of sets \cite{halmos}\\

\hspace{6cm} $\textit{L} \bigotimes
 \textit{M} \bigotimes \textit{C}$\\
we can be assured that any statement regarding the observable
world is contained as a true statement p(z) contained in a set

 \hspace{3cm}$\textsf{P} = \{p(z)|z\in \textit{L}\bigotimes
\textit{M} \bigotimes  \textit{C}\hspace{0.1cm} s.t
 \hspace{0.1cm}p(z)
\hspace{0.1cm} is \hspace{0.1cm}true\}$\\

 So the maximum information accessible to
any observer is encoded in the set $\textsf{P}$. It is the set of
objects which are knowable to any observer. Hence in order to
discuss the discovery of physical theories it is quite necessary
to study the possible evolutions of the set . How can such a
possible evolution be achieved is a dynamical question also
completely dependent on the structures of the set $\textsf{P}$,
which at this stage is pretty difficult to answer. We on the other
hand will be mainly concerned with the mechanical model of what
may be the possible sorts of evolution.

Borrowing our notations from Dirac's notation. we claim that the
set $\textsf{P}$ corresponding to the observable world is
associated a state $|\textsc{P}\rangle$. The state
$|\textsc{P}\rangle$ encodes the information about the observed
world. Now as we have got a glimpse of the basic scheme we now
state our basic
propositions.\\

\textbf{Proposition 1:}\\
The observed states is the effect of the interaction between the
observer (brain states) \cite{neuro}and the real world states. The
interaction is in particular a form of entanglement,in a sense
that the real world states may influence the observer (brain)
states and therefore correlated .{\footnote{ For notational
convenience we term the Observer + apparatus as the brain states.
The symbolism should not be misunderstood with the connotation of
associated consciousness.}}
\\
So symbolically,\\

\hspace{6cm}
 ${|\textsc{P}\rangle} \equiv \sum_{k} c_{k}{{|\textsc
B}_{k}\rangle} {{|\textsc R}_{k}\rangle}$ \footnote{ For
simplicity of notations from now onwards we don't include the sum
though we represent an entangled state.}

\textbf{Proposition 2:}\\
Corresponding to $i\in I$ where $I$ is
 an indexing set,there exists a class
 of states
${|{\textsc B}_{i}\rangle}
 {|{\textsc R}_{i}\rangle}$
which correspond to different class of
 observed states given by
 ${|{ \textsc P}_{\lambda}\rangle}$ where
 $\lambda \in \Delta$ where $\Delta$
 is a different indexing
 set.

\textbf{Proposition 3:}\\

Reaching Conclusions from a given state ${{|\textsc
P}_{i}\rangle}$ will be induced by at most two inequivalent
processes , which are marked as \emph{X} and \emph{O}. The
process \emph{X} denotes a computational process and the process
\emph{O} which is not computational. At this stage the terms
computational and non computational are defined ostensively.

\textbf{Proposition 4:}\\

The brain states${|\textsc{B}\rangle}$ are associated to the
observer and do correspond to the brain functions which is
essentially the seat for undergoing computational processes.\\

Now to describe the evolution of the state ${|{ \textsc
P}_{\lambda}\rangle} = {|{\textsc B}_{i}\rangle}
 {|{\textsc R}_{i}\rangle}$ to a completely different non trivial
state ${|{\widetilde{\textsc P}}_{\lambda}\rangle} =
{{|\widetilde{\textsc B}}_{i}\rangle} {{|\widetilde{\textsc
R}}_{i}\rangle}$ is denoted
 schematically by \\

 \hspace{5cm} ${|{\textsc P}_{\lambda}\rangle}$
 $\stackrel{\emph X \hspace{.05cm}\& \hspace{.05cm}\emph O }
 {\longrightarrow}$${|{\widetilde{\textsc
P}}_{\lambda}\rangle}$

\textbf{Proposition 5:}\\

The Gedanken experiment which we performed gives rise to another
conclusion which we state as a proposition. Same set of real
world states may give rise to different observable states for the
same brain states. Different sets of real world states may also correspond to
different observed states for the same set of brain states.\\
Symbolically,\\
\begin{center}
${|{\textsc P}_{\lambda}\rangle} = {|{\textsc B}_{i}\rangle}
 {|{\textsc R}_{i}\rangle}$ \hspace{2cm}
 ${|{\textsc P}_{\lambda}^{'}\rangle} = {|{\textsc B}_{i}\rangle}
 {|{\textsc R}_{i}^{'}\rangle}$\\
\end{center}
\vspace{1cm}
 This is a consequence of the third proposition which
entails that any conclusion whatsoever from a definite
information state corresponding to the observed world evolves
into any other information state by 2 fundamental inequivalent
processes. At this point it is imperative to attend to the basic
problem of measurement as it is in quantum mechanics.As it is
clear that we have associated the basic knowledge about the
observed world as a class of entangled  information states. So
essentially any gain in information is but an evolution of the
associated state. Though we will propose a model for the
evolution but the nature is not yet clear. If it is a state
collapse then the role of observer is pretty ambiguous.For some
alternative proposals London $\&$ Bauer \cite{london} has an
interesting suggestion,though they unlike us do not have a
construction of the real world states. According to their
proposals that states of the observer satisfy vectorial relations
required by ordinary quantum mechanics,but do not evolve
temporally in the quantum mechanical sense.Though in a sense it
is a strange proposal,extending quantum theory somewhat into the
domain of psychology,nevertheless a consistent theory.In other
words the proposal claims an inclusion of the observer to
be somewhat essential for the state vector collapse.\\

Another pretty different but interesting idea stems from Von
Neumann \cite{von} \\

 \begin{quotation}
  \textit{the so-called principle of psycho-physical parallelism -that it
  must be possible so to describe the extra physical process of
  the subjective perception as it were in reality in the physical
  world-i.e, to assign to its parts equivalent processes in the
  objective environment,in physical space.}
\end{quotation}

In accordance with this argument whenever there is a measurement,
the world is divided into two parts, one part comprising the
system being observed and the other the observer. The boundary
between these parts is arbitrary. According to Von Neumann if the
formalism of quantum mechanics is applied to the observed part of
the world, the predictions obtained are independent of the
location of the boundary.

 At present we will not analyze the
above issues any furthur. Our point has been to state how to
model an observer in our context and we argue that there may be
various alternative approaches which may be undertaken but the
issue is very well unsettled at present.

 Equipped with these
propositions let us now try to find out in detail how to obtain
conclusions from a set of premises. As it is obvious the set of
premises do form a definite state obtained from the set P.

 Any
conclusion is thereby defined to be the implementation of a
series of states
\begin{center}
 ${|{\textsc P}_{\lambda}\rangle}
{\longrightarrow}{|{\widetilde{\textsc P}}_{\lambda}^{1}\rangle}
{\longrightarrow}{|{\widetilde{\textsc P}}_{\lambda}^{2}\rangle}
{\longrightarrow} \cdots$
\end{center}

 Now say we start with an
information which corresponds to an observed state

\begin{center}
${|{\textsc P}_{\lambda}\rangle} = {|{\textsc B}_{i}\rangle}
 {|{\textsc R}_{i}\rangle}$
\end{center}

 By Proposition 4,the ${{|\textsc B}_{i}\rangle}$ states
transform by the computational process {\emph X} through a series
(the accessible states)

\begin{center}
 ${|{\textsc B}_{i}\rangle}
 \stackrel{\emph X}
 {\longrightarrow}{|{\textsc
B}_{i}^{'}\rangle}$
\end{center}

This may in turn give rise to a new observed state

\begin{center}
${|{\textsc P}_{\lambda}^{'}\rangle} = {|{\textsc
B}_{i}^{'}\rangle}
 {|{\textsc R}_{i}\rangle}$
\end{center}

So preceding as above we may get a series of states

\begin{center}
 ${|{\textsc P}_{\lambda}\rangle}
{\longrightarrow}{|{\textsc P}_{\lambda}^{1}\rangle}
{\longrightarrow}{|{\textsc P}_{\lambda}^{2}\rangle}
{\longrightarrow} \cdots  {|{\textsc P}_{\lambda}^{n}\rangle}$
\end{center}

which are connected with each other by the transformation of the
brain states via a computational process.

At this stage we link th other process which is fundamentally
different from the computational one in a sense that they are
nonlocal and stochastic in nature and denote a jump of the states
which may not be determined. In this context it will be
interesting to point out the question of the quantum jumps has
been previously being considered by \cite{werner}.The process is
marked by the special property that the real world states may
influence the brain states to deliver a jump.

\begin{center}
 ${|{\textsc P}_{\lambda}\rangle}
\equiv {|{\textsc B}_{i}\rangle}{|{\textsc R}_{i}\rangle}
\stackrel{\emph X}{\longrightarrow}{|{\textsc B}_{i}^{'}\rangle}
{|{\textsc R}_{i}\rangle}{\longrightarrow} {|{\textsc
B}_{i}^{'}\rangle}{|{\textsc R}_{i}^{'}\rangle}$
\end{center}

The last step of the transformation has been in accordance with
the Fifth Proposition which entails that there exists states for
the same brain states but different real world states.

The next process which we show is another assumption on the basis
of Proposition 3 where we give a concrete scheme for
non-computational process.

\begin{center}
${|{\textsc B}_{i}\rangle}{|{\textsc
R}_{i}^{'}\rangle}\stackrel{\emph O}{\longrightarrow}{|{\textsc
 {\"{B}}}_{i}^{o}\rangle}{|{\textsc R}_{i}^{'}\rangle}$
\end{center}

Thereby the ${|{\textsc
 {\"{B} }}_{i}^{o}\rangle}$
 are not accessible from the ${|{\textsc
 B}_{i}\rangle}$ state by any computational process.

So now we conclude that any sort of nontrivial conclusions is
being generated by the following sequence

\begin{center}
 ${|{\textsc P}_{\lambda}\rangle}
\equiv {|{\textsc B}_{i}\rangle}{|{\textsc R}_{i}\rangle}
\stackrel{\emph X}{\rightarrow}{|{\textsc
B}_{i}^{1}\rangle}{|{\textsc R}_{i}\rangle}\stackrel{\emph
X}{\rightarrow}{|{\textsc B}_{i}^{2}\rangle}{|{\textsc
R}_{i}\rangle}\equiv{|{\textsc B}_{i}^{2}\rangle}{|{\textsc
R}_{i}^{'}\rangle}\stackrel{\emph O}{\rightarrow}{|{\textsc
 {\"{B}}}_{i}\rangle}{|{\textsc R}_{i}^{'}\rangle}
 \stackrel{\emph X}{\rightarrow}
{|{\textsc {\"{B}}}_{i}^{1}\rangle}{|{\textsc R}_{i}^{'}\rangle}
 {\rightarrow}
{|{\textsc {\"{B}}}_{i}^{1}\rangle}{|{\textsc R}_{i}^{''}\rangle}$
\end{center}

So the state $ {|{\textsc {\"{B}}}_{i}^{1}\rangle}{|{\textsc
R}_{i}^{''}\rangle} $ can be identified with the non trivial final
state ${|{\textsc {\"{P}}}_{\lambda}\rangle}$. The states are
generated by a series of computational $\&$ non-computational
processes. We propose that for gaining information which is not
encoded in the initial state the computational processes are not
enough, though we have assumed that he brain states are connected
by those only.

Now we give an example from computational geometry
\cite{discrete}illustrating our scheme. We took this particular
example in particular to show that though we may be inclined to
think that this is wholly a computational problem but essentially
we show here that how can our scheme may be implemented in this
case and this in turn also shows how noncomputational processes
play pivotal role in the theory of inferences.

In essence computational geometry is the study of designing and
analysis of algorithms in the context of problems from
geometry.In view of the above analysis we would like to see how
the above mechanism of generating conclusions translate to the
case of closest pair problem.\\


\textbf{The Premise:}\\
Let there are given points in a plane. What are conclusions which
we can reach from these set of data.

\textbf{The Analysis:}\\
We will be mainly concerned with the simplest solution for the
above problem in accordance with the model proposed by us.

As the premise consists of given points on a plane it is the
concerned observed state ${|{\textsc {P}}_{\lambda}\rangle}$. As
we have already discussed this state is an entangled state of the
corresponding real world and brain states. So,
\begin{center}
${|{\textsc {P}}_{\lambda}\rangle} = \sum_{i} c_{i}{|{\textsc
{B}}_{i}\rangle}{|{\textsc R}_{i}\rangle}${\footnote{We made an
exception here with respect to our earlier remarks. Again we state
that for simplicity from now onwards we don't include the sum
though we represent an entangled state.}}
\end{center}
So qualitatively the points are designated as an entangled state
corresponding to some real world which cannot be determined and
the brain world states.

So to get some non trivial conclusions from the initial state
${|{\textsc {P}}_{\lambda}\rangle}$ we get at first the following
sequence.

\begin{center}
 ${|{\textsc P}_{\lambda}\rangle}
\equiv {|{\textsc B}_{i}\rangle}{|{\textsc R}_{i}\rangle}
\stackrel{\emph O}{\rightarrow}{|{\textsc
 {\"{B}}}_{i}\rangle}{|{\textsc R}_{i}^{'}\rangle}
\equiv{|{\textsc P}_{\lambda}\rangle}$
\end{center}
So here a non computational process which transforms the brain
state ${|{\textsc B}_{i}\rangle}$ into a brain state ${|{\textsc
 {\"{B}}}_{i}\rangle}$ which is the state which registers the
concept of a distance between the points in the plane or may be
also the number of points in the plane. It should be noted here
that this is already a complex process which has other parameters
too. In order to give a simplified model for the aforesaid scheme
we keep the complexities to a minimum. So in brief the
noncomputational process gives rise to a state which may now
effectively give rise to some non trivial conclusions which are
not encoded in the initial state which may be described as
follows.

The transformed brain state ${|{\textsc
 {\"{B}}}_{i}\rangle}$ in this case of the simplified model is the
 seat for generating nontrivial computational processes
which we may depict as \\
\begin{center}
${|{\textsc P}_{\lambda}\rangle} \equiv {|{\textsc
 {\"{B}}}_{i}\rangle}{|{\textsc R}_{i}^{'}\rangle} \stackrel{\emph
 X}\Longrightarrow {|{\textsc P}_{\lambda}^{'}\rangle}$
\end{center}

So essentially as with the closest pair problem,state which has
been obtained by a noncomputational process now gives rise to a
state which via some computational processes only give rise to
some entangled real and brain world states which corresponds to
some observed state.  The computational process as it can be
guessed can be generated by an algorithm. We are not interested
here to analyze the efficiency of the algorithm

\section{Conclusion}

So as a matter of fact the main objective of this paper has been
to go through in detail the process by which we may make definite
conclusions which has been speculative and we proposed a model as
regards to it. The accuracy of the model will nevertheless be
dependent on correct theories of computation, quantum aspects and
neural networks. The theory of cognitive neuroscience
\cite{neurobio}and computational neurobiology
\cite{neurobiol}will be immensely helpful in deciphering neural
network models which may be in turn be focused to predict a model
for cognitive function. The problem which we feel here is that in
computational correlates of brain function and associated aspects
models may be constructed at various stages which may be
consistent with each other. For example the Hopfield
\cite{hopfield}neural network model along with back propagation
and classical theory of computation may be used to give rise a
specific model for perception. In the context of
information-processing approaches mental or cognitive events are
abstract and cannot be reduced to neurological phenomena.
Perception, memory learning etc may be mapped in terms of their
functions and processes without concern for their neurological
causes. Though at this point it will be worthwhile to ask whether
neural activities of brain can be modeled by quantum mechanics
but some recent arguments more or less tries to answer the
scepticism in the negative \cite{1}\cite{2}\cite{3}\cite{4} .As
far as the computational viewpoint is concerned a logical theory
machine \cite{logic1}has been constructed for the first
mathematical proof conducted by a machine,there are also claims
of General Problem Solver to study and simulate human problem
solving. On the extreme of these viewpoints there has been
connectionist \cite{conn}approaches which has close ties to the
neurological perspective that the mind is powered by the brain
where a serial processing of information sequential, or in
parallel, processes in the reception and transformation of
information signals. In this context it will be interesting to
remark at this stage that for a specific model of memory we
believe that there exists some optimal physical process, though
we feel it will be quite remarkable if something like Grover's
algorithm \cite{grover} has any role to play there. We hope to
construct a category based model of perception and thought in our
next endeavor.

The model we have proposed here essentially underpinnes a
noncomputational process which we at this stage have defined
ostensively. But we believe that some appropriate neurobiological
process should be accounted for that. As a remark it is felt that
the sacred laws of physics which govern nature may not be
universal but relative to the cognitive system. As an example two
different neural models may give rise to completely different
perception and modes of inference processing on account which
they both may differ in their corresponding world views.

\section*{Acknowledgements:}

I.Mitra would like to acknowledge the hospitality of Theory
Group,Saha Institute Of Nuclear Physics, and Indian Statistical
Institute, Kolkata for this work.


\end{document}